\documentclass[%
 reprint,
 amsmath,amssymb,
 aps,
]{revtex4-2}

\usepackage{graphicx}
\usepackage{dcolumn}
\usepackage{bm}
\usepackage{hyperref}

\begin{document}


\title{Switching current distributions in superconducting nanostrips}


\author{Robert Vedin}
\email{rvedin@kth.se}
\author{Jack Lidmar}%
\email{jlidmar@kth.se}
\affiliation{%
 Department of Physics, KTH Royal Institute of Technology, SE-106 91 Stockholm, Sweden
}%


\date{\today}

\begin{abstract}
We study switching current distributions in superconducting nanostrips using theoretical models and numerical simulations.
Switching current distributions are commonly measured in experiments and may provide a window into the microscopic switching mechanisms.
As the current through a superconducting strip is increased from zero it will at some point switch to the normal dissipative state.
Due to thermal and/or quantum fluctuations the switching current will be random and follow a certain distribution depending on sweep rate, temperature, material properties and geometry.
By analyzing the resulting distribution it is possible to infer the transition rate for a switch, which can be related to the free energy barrier separating the metastable superconducting state and the normal one.
We study different switching scenarios and show using simulations how data taken for different sweep rates can be combined to obtain the switching rate over a wider interval of currents.
\end{abstract}

\maketitle

\section{\label{sec:intro} Introduction}

Superconducting nanowires have emerged as an important component in applications such as the superconducting nanowire single photon detector (SNSPD) \cite{Natarajan2012}.
These devices rely on a bias current close to the critical $I_c$ creating a fragile metastable superconducting state, such that the perturbation of a single photon is sufficient to trigger a switch to the normal state.
In this bias regime the detectors also become sensitive to random fluctuations that can cause breakdown of the superconductivity in the form of dark counts.

Dark counts may be triggered by thermally activated phase slips in thin superconducting wires~\cite{Langer1967,McCumber1970b}
or quantum phase slips if the temperature is low enough~\cite{Aref2012,Murphy2015}.
In superconducting nanostrips a phase slip involves the entry, or unbinding, of vortices and anti-vortices~\cite{AHNS1980,Engel2006,Kitaygorsky2007}.
The rate of these thermally activated events is often described by an Arrhenius type law governed by a free energy barrier. 
The problem of calculating this energy barrier has previously been approached using different methods~\cite{Langer1967,McCumber1970b}.
Analytical estimates have been found by considering the interaction energy of mirror vortices in strips~\cite{Gurevich2008, Bulaevskii2011}.
Numerical works based on the string method have also been demonstrated \cite{ChunyinQiu2008, Benfenati2020}, however the application of this method has been limited to consider only cases with no bias current.
Similar mechanisms that are responsible for the dark counts also give rise to random variations of the switching current. The switching current statistics could therefore be used as an additional method of extracting information about the vortex energy barrier in the current biased regime.

In this work the switching current distribution due to thermal activation is investigated theoretically and numerically through stochastic time dynamics for two different models of a superconducting wire, a one-dimensional (1D) Josephson junction chain and a two-dimensional (2D) time-dependent Ginzburg-Landau model.
The switching current distribution in homogeneous wires is here shown to depend on the sweep rate of the bias current in a way that is analogous to the wire length.
This behaviour can be exploited in order to obtain the switching current statistics, as well as the switching rate, on a larger interval of currents.
In particular, this can be relevant for experiments, where changing the length of a wire typically involves fabrication of a completely new device, which can introduce additional variation of the $I_{sw}$ due to inhomogeneities arising from the fabrication process.
The length dependence of such sample-to-sample variations due to material disorder and inhomogeneities has previously been studied experimentally~\cite{Gaudio2014, XiaoyanYang2018} and theoretically~\cite{Chen2024} in absence of thermal fluctuations.

In Sec.~\ref{sec:theory} we discuss theoretical approaches to compute the switching rate and arguments connecting it to the distribution of switching currents.
Section~\ref{sec:examples} illustrates the approach using several relevant examples from different switching scenarios.
Section~\ref{sec:TDGL} describes a larger scale numerical simulation based on time-dependent Ginzburg-Landau theory.

\section{Theory}
\label{sec:theory}

A current carrying superconducting wire will be in a metastable state where  typically a large free energy barrier protects the current from decaying.
As the applied current is increased the barrier will become lower and eventually vanish at some critical current $I_c$.
In thin wires, thermal (or quantum) fluctuations can, however, cause phase slips and a decay of the current even below the critical current.
This in principle results in a finite resistance $R \sim e^{-\Delta U/k_B T}$, at any finite temperature, although it will be exponentially suppressed when the barrier is large compared to temperature.

We will focus on the situation where thermal fluctuations dominate over quantum, and where the applied current is relatively large but below $I_c$.
This means that once a fluctuation is large enough to initiate a phase slip at some location in the wire, enough energy is released to locally cause a transition to the normal state, which may be detected as a voltage along the wire.

The rate $\Gamma$ of such switching events is mostly controlled by an energy barrier $\Delta U(I)$, which for a given geometry depends on the applied current $I$.

\subsection{Switching current distributions}
\label{sec:switching-current-distr}

Consider an applied current $I(t)$ gradually increasing from $0$ at $t = 0$ to some value above the nominal critical current $I_c$.
We will initially assume that the switching rate $\Gamma_L(t) \equiv L \Gamma(t)$ of a homogeneous wire of length $L$ depends on the applied current, but not on the sweep rate $\dot I \equiv dI/dt$, i.e., $\Gamma_L = L \Gamma(I(t))$, which may be expected to hold for small enough $\dot I$.
Under these circumstances it is possible to relate the switching rate to the probability distribution of the switching current, following arguments pioneered by Kurkijärvi and Fulton-Dunkleberger~\cite{Kurkijarvi1972,Fulton1974}.
Assuming that switching may occur independently for each infinitesimal time interval $\Delta t$, the probability of having no switch during a time 0 to $t = n\Delta t$ will be
\begin{equation}
    S(t) = \prod_{i = 1}^n \left( 1 - \Gamma_L(I(i \Delta t)) \Delta t \right) .
\end{equation}
In the limit $\Delta t \to 0$ this probability becomes
\begin{equation}
    S(t) = e^{-\int_0^t \Gamma_L(t') dt'} .
\end{equation}
The switching rate $\Gamma_L(I)$ may then be obtained as
\begin{equation}
    \Gamma_L(I(t)) = - \frac{\partial}{\partial t} \ln S(t).
\end{equation}
Specializing in the following to the case where the applied current is ramped up linearly from zero,
\begin{equation}
    I(t) = \dot I t,
\end{equation}
with a constant sweep rate $\dot I$, and assuming a homogeneous wire where the switching event may occur equally likely along the whole length $L$, we may write
the cumulative distribution $F(I) = \Pr(I_\mathrm{sw} < I) = \int_0^I P(I') dI' = 1 - S(I/\dot{I})$ for the switching current $I_\mathrm{sw}$ as
\begin{equation}                                            \label{eq:cummulative}
    F(I) = 1 - \exp\left({ - \frac{L}{\dot I} \int_0^I \Gamma(I') dI'}\right),
\end{equation}
where $\Gamma = \Gamma_L /L$ is the switching probability per time and length. The latter may thus be extracted via
\begin{equation} \label{eq:Gamma_from_cumulative_dist}
    \Gamma = - \frac{\dot I}{L} \frac{\partial}{\partial I} \ln \left( 1 - F(I) \right) = \frac{\dot I}{L} \frac{P(I)}{1 - F(I)}
\end{equation}
from measured (or simulated) switching current distributions.
According to Eq.~\eqref{eq:cummulative} the sweep rate and the length of the wire only enter in the combination $L/\dot I$, hence the effect of increasing the length of the wire will be the same as decreasing the sweep rate.

\subsection{Delay time}                                                     \label{sec:delay}

It turns out that the picture presented in the previous section may need some modifications when compared to numerical simulations or experiments.
The assumption that the switching rate $\Gamma$ does not depend on the current sweep rate $\dot I$ is only approximately valid.
In particular, there may be a time delay $\tau_d$ between the initiation of a switching event and the detection of this event.
During this delay-time the current will increase further, which will induce a systematic shift in the distributions that must be accounted for.
Experimentally, the delay may be due to the time of propagation of the voltage pulse through the wire to the detector, but there can also be an intrinsic delay in the nucleation mechanism of a phase slip, as discussed in more detail below.

The probability distributions $F(I)$ and $P(I) = \partial F(I)/\partial I$ of the previous section then correspond to the initiation of a switching event. The detection will occur after a slight delay $\tau_d$, during which the current has increased by $\int_t^{t+\tau_d} \dot I dt \approx \dot I \tau_d$.
In the simplest setting we may assume that the delay time $\tau_d$ is constant, independent of $I$ and $\dot I$.
The cumulative distribution for the detection is then $F^\mathrm{det}(I,\dot I) = F(I - \dot I \tau_d)$, and depends also on the sweep rate $\dot I$.
This complicates the analysis, since the transition rate $\Gamma(I)$ can no longer be extracted from Eq.~\eqref{eq:Gamma_from_cumulative_dist} using $F^\mathrm{det}$ in place of $F$.
More generally, we may assume that the delay time is random with a certain distribution $P_{\tau_d}(\tau_d)$, so that
\begin{equation}
    F^\mathrm{det}(I, \dot I) = \int\limits_0^\infty P_{\tau_d}(\tau_d)
    F(I - \dot{I} \tau_d)
    d\tau_d .
\end{equation}
In this case we may still define a characteristic delay time $\tau_d^*(I,\dot I)$ so that
$F^\mathrm{det}(I,\dot I) = F(I - \dot I \tau_d^*)$, and derive an effective
$\Gamma^\mathrm{eff}(I,\dot I) = - \frac{\dot I}{L} \frac{\partial}{\partial I}\ln (1 - F^\mathrm{det}(I,\dot I)) = (1 - \dot I \partial \tau_d^*/ \partial I) \Gamma(I - \dot I \tau_d^*)$, related to the true one by a shift and a scale factor.
If the distribution of $\tau_d$ is very narrow  $\tau_d^*$ will only be weakly dependent on $I$ and may be treated as constant.
We will test this hypothesis in simulations below.

\subsection{Escape over barrier}

To make a more detailed study a microscopic model of the switching mechanism is needed. We will discuss several such models in this and the following sections.
The most important parameter determining the switching probability is the energy barrier $\Delta U(I)$.
When the temperature is low compared to $\Delta U$ the rate will follow an Arrhenius law
\begin{equation}        \label{eq:rate}
    \Gamma \approx \frac{\Omega(I)}{2\pi}\, e^{-\Delta U(I)/k_B T} .
\end{equation}
Although the dynamics of the transition typically involves a large number of degrees of freedom, it is often possible to project it down to a single \emph{reaction} coordinate $x$, starting at $x_0=0$ in the uniformly superconducting metastable state and reaching a final value $x_1$ on the other side of the (free) energy barrier $\Delta U(x,I)$.
The maximum barrier $\Delta U(I) = \max\limits_x \Delta U(x,I)$ occurs at some $x_{\max}$ given by $\partial \Delta U(x_{\max},I)/ \partial x = 0$, and will decrease with increasing current until it becomes zero at the nominal critical current.
The prefactor $\Omega/2\pi$ can be estimated using Kramers' theory, as
$\Omega = \omega_{\max} \omega_{\min} D/k_B T$,
where $\omega_i^2 = |\partial^2 \Delta U(x_i,I) / \partial x_i^2|$ evaluated at the minimum $x_i = x_{\min}$ and the maximum $x_{\max}$, and $D$ is an effective diffusion constant along $x$.

Alternatively, instead of relying on the approximate Eq.~\eqref{eq:rate} the rate can be obtained from the mean first passage time.
In the following, $x(t)$ is assumed to obey a one-dimensional overdamped Langevin equation, $\dot x = -D \beta \Delta U'(x) + \zeta(t)$, where $D = k_B T/\alpha$ is the diffusion constant, $\beta = 1/k_B T$, and $\zeta$ is a white noise process with zero mean and covariance $\left< \zeta(t) \zeta(t') \right> = 2D \delta(t-t')$.
The applied current is held fixed in this argument and we write $\Delta U(x,I)=\Delta U(x)$ for brevity.
The corresponding Fokker-Planck equation for the probability $P(x,t)$ is
\begin{equation}
    \frac{\partial}{\partial t} P(x,t) = \frac{\partial}{\partial x} D e^{-\beta \Delta U(x)}
    \frac{\partial}{\partial x} e^{\beta \Delta U(x)} P(x,t) .
\end{equation}
The transition rate $\Gamma$ may then be directly related to the mean first passage time $\tau$ for crossing the barrier via $\Gamma = \tau^{-1}$.
The latter may be expressed in closed form as~\cite{Gardiner}
\begin{equation}                \label{eq:MFPT}
    \tau = \frac{1}{D} \int_{x_0}^{x_1} dx e^{\beta \Delta U(x)} \int_{x_0}^{x} dy e^{-\beta \Delta U(y)} ,
\end{equation}
which may be evaluated numerically to give the transition rate also when the condition $\Delta U(I) \gg k_B T$ does not hold.
In the following sections we consider a few different scenarios for the switching mechanism.

\section{Illustrative examples}
\label{sec:examples}

\subsection{Single Josephson junctions}
\label{sec:JJ}

As a first illustration consider a single Josephson junction connected to a current source and shunted by a resistance $R$.
The reaction coordinate $x$ may here be identified with the phase difference $\phi$ of the superconducting order parameter across the junction.
In the overdamped limit, i.e., neglecting the junction capacitance, the phase obeys a Langevin equation
$(\Phi_0/2\pi R) \dot \phi = - I_c \sin\phi + I + I_n(t)$
where $I_n$ is the Johnson-Nyquist noise in the resistor with zero mean and $\left<I_n(t)I_n(t')\right> = (2 k_B T /R) \delta(t-t')$.
This leads to an effective phase diffusion constant $D =(2\pi/\Phi_0)^2 R k_B T$.

\begin{figure}
    \centering
    \includegraphics[width=1\linewidth]{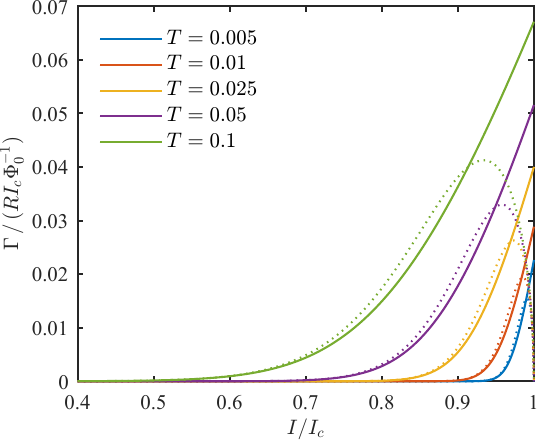}
    \caption{Comparison of the switching rate $\Gamma$ computed from the mean first passage time (solid lines) and using Kramers' theory (dashed lines) for a single Josephson junction. The temperature is measured in units of $E_J$. }
    \label{fig:JJrate}
\end{figure}

The corresponding energy as function of $\phi$ takes the form of a tilted washboard potential $\Delta U(\phi,I) = - E_J \cos \phi - (\Phi_0 / 2\pi) I \phi$, where $E_J = I_c \Phi_0/2\pi$.
When the applied current is below the critical, $I < I_c$, the stationary solution yields $\phi_{\min} = \sin^{-1} (I/I_c)$, while the barrier maximum occurs at $\phi_{\max} = \pi-\phi_{\min}$.
The energy barrier becomes $\Delta U(I) = 2 E_J \sqrt{1 - (I/I_c)^2} + I \Phi_0(\pi - 2 \phi_{\min})/2\pi$, and
$\omega_{\max} = \omega_{\min} = \sqrt{E_J\cos \phi_{\min}} = \sqrt{ (\Phi_0/2\pi) \sqrt{I_c^2 - I^2} }$, so that the transition rate estimated using Kramers' theory will be~\cite{Ambegaokar1969}
\begin{equation}                \label{eq:JJrate}
    \Gamma(I) =    
    \frac{R \sqrt{I_c^2 - I^2}}{\Phi_0} e^{-\Delta U(I)/k_B T} .
\end{equation}
In Fig.~\ref{fig:JJrate} we compare this with the more precise $\Gamma = \tau^{-1}$ obtained by numerically integrating Eq.~\eqref{eq:MFPT}. As seen the Kramers rate will be relatively accurate except when the barrier is low compared to temperature. The downturn at high currents of the latter approximation comes from the prefactor and is obviously unphysical.

Below we will compare with the value extracted from simulated switching current distributions in Josephson junction chains.

\subsection{Josephson junction chains}

As a more complex test case,
we consider simulated switching current distributions in a chain of $N$ Josephson junctions connected in series.
We employ the circuit model described in Refs.~\cite{Ergul2013,Ergul2017}.
Each junction of the chain is modeled as an ideal Josephson junction 
shunted by a capacitance $C$ and a nonlinear resistor $R$.
The total current through junction $i$ is
\begin{align}\label{eq:1}
  I^\mathrm{tot}_i &= I^s_i + I^C_i + I^R_i \nonumber \\
&= 
I_{c} \sin(\theta_{i} - \theta_{i+1}) + C (\dot V_{i} - \dot V_{i+1}) + I^R_i ,
\end{align}
where $\theta_i$ is the phase of the superconducting order parameter
at the island to the left of junction $i$, and
$V_i =  \dot \theta_i \Phi_0 / 2\pi$
is the voltage.
The nonlinear resistive current is taken to be
\begin{equation}							\label{eq:nonlinear}
  I^R_i =
  \begin{cases}
    (V_{i} - V_{i+1})/R + I^n_i & \text{if}\,  |V_{i} - V_{i+1}| > V_g \\
    (V_{i} - V_{i+1})/R_\mathrm{qp} + I_i^{\mathrm{qp},n} & \text{otherwise}
  \end{cases},
\end{equation}
where $R$ is the normal resistance of a single junction, and $R_\mathrm{qp} \gg R$ is the quasiparticle subgap resistance.
The current entering the chain through the lead resistance is given by
\begin{equation}
  \label{eq:3}
  I_L = (U - V_1)/R_\text{term} + I^n_L ,
\end{equation}
where $U$ is an applied voltage. In addition thermal noise currents $I_i^n$ are included.
They are modeled as a Gaussian random Johnson-Nyquist noise with zero mean and covariance
$\left< I_i^n(t) I_j^n(t')\right>
= (2 k_B T/R_i) \delta_{ij} \delta(t-t')$.
The dynamical equations of motion form a system of equations obtained by imposing current conservation at each island $i$,
\begin{equation}
  \label{eq:4}
  C_0 \dot V_i + I^\text{tot}_{i} - I^\text{tot}_{i-1} = 0,
\end{equation}
where $C_0$ is the capacitance to ground.
This gives a coupled system of 2nd order differential equations for
the superconducting phases $\theta_i$.
These are integrated using a symmetric time discretization with a small time step
$\Delta t = 0.02 (\Phi_0/2\pi I_c R)$.
Each iteration requires the solution of a tridiagonal system of equations.

We set $E_J/E_C = (I_c\Phi_0/2\pi)/(4e^2/C) = 1$, $C/C_0 = 100$, $R_{qp}/R = 100$, $T = 0.01 E_J$, $R_\text{term} = 200 R$, $V_g = R I_c$, and vary the sweep rate $\dot I$ from $10^{-7}$ to $10^{-3}$ in units of $2\pi R I_c^2/\Phi_0$, for a chain consisting of $L = 100$ junctions.
For this parameter choice with $C \gg C_0$, the phase slips will be highly localized to single junctions and the reaction coordinate can be defined as the superconducting phase difference $\phi = \Delta \theta$ across a junction in accordance with Sec.~\ref{sec:JJ}.
Furthermore, a single phase slip at a particular junction will cause it to latch and stay in the dissipative running state, so a switching event may be identified as the first phase slip event.

\begin{figure}
\includegraphics[width=1 \linewidth]{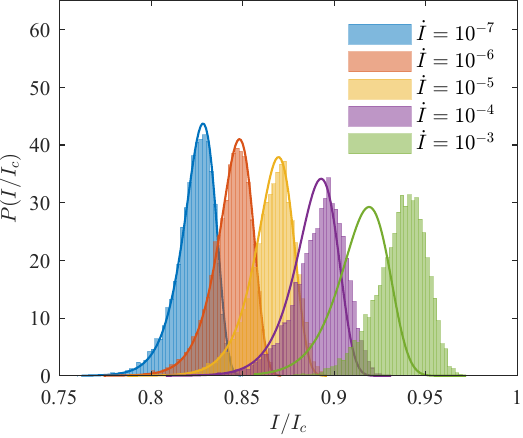}%
\caption{\label{fig:JJ}%
Switching current distributions obtained from simulations of the Josephson junction chain using 10 000 realizations for each different sweep rate $\dot I$. The solid lines show the corresponding prediction from the mean first passage time numerically computed from Eq.~\eqref{eq:MFPT}. The temperature was set to $0.01 E_J$.}
\end{figure}

We show in Fig.~\ref{fig:JJ} the resulting simulated switching current histograms together with the theoretical prediction (solid lines) obtained from Eq.~\eqref{eq:cummulative} using the rate $\Gamma = \tau^{-1}$ numerically computed from Eq.~\eqref{eq:MFPT}. For low sweep rates the agreement is very good, considering that no fitting parameters were adjusted, in spite of the Josephson junction chain being considerably more complicated than a single junction. For higher sweep rates deviations are clearly seen, presumably because the time delay discussed in Sec.~\ref{sec:delay} becomes non-negligible.
The differences in the predictions of the distributions from Kramers' theory, Eq.~\eqref{eq:JJrate} and the mean first passage time are minuscule in this case.

\begin{figure}
    \centering
    \includegraphics[width=1\linewidth]{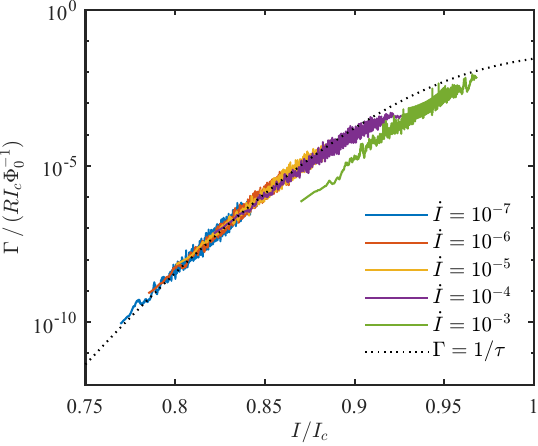}
    \caption{Switching rate $\Gamma$ recovered from the switching current histograms in Fig.~\ref{fig:JJ} using Eq.~\eqref{eq:Gamma_from_cumulative_dist} for the Josephson junction chain. The dashed line shows the inverse mean first passage time computed from Eq.~\eqref{eq:MFPT} for comparison.}
    \label{fig:JJsimrate}
\end{figure}

From the empirical histograms it is possible to recover the switching rate from Eq.~\eqref{eq:Gamma_from_cumulative_dist}. This is shown in Fig.~\ref{fig:JJsimrate}.
At least for low sweep rates the accuracy of the procedure appears satisfactory.

\subsection{One-dimensional time-dependent Ginzburg-Landau theory}

\begin{figure}
    \centering
    \includegraphics[width=1\linewidth]{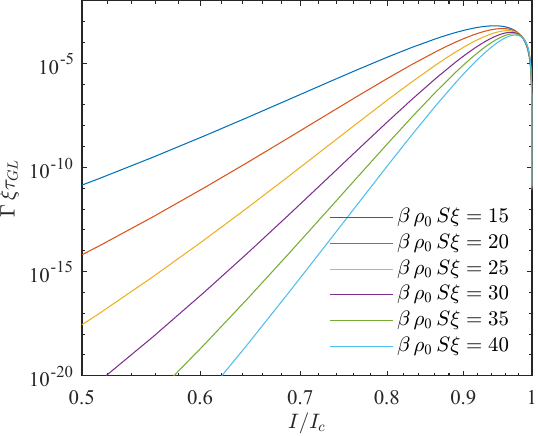}
    \caption{Switching rate $\Gamma$ from the LAMH theory for a one-dimensional superconducting wire. The different curves corresponds to different temperatures.}
    \label{fig:TDGL-rate}
\end{figure}

Thin continuous wires are more appropriately modeled by time-dependent Ginzburg-Landau theory.
Within a one-dimensional TDGL description, the free energy barrier for thermal phase slips has been calculated by Langer and Ambegaokar and by McCumber and Halperin (LAMH) as~\cite{Langer1967,McCumber1970b}
\begin{align}            \label{eq:LAMH}
    \frac{\Delta U(\kappa)}{\rho_0 S \xi} &= \frac{8\sqrt{2}}{3} \sqrt{1-3 \kappa^2 }
    \nonumber \\
    & - 8\kappa {\left( 1 - \kappa^2 \right)} \tan^{-1}\left(\frac{\sqrt{1-3\,\kappa^2}}{ 2 \kappa }\right)
\end{align}
where $\kappa$ is related to the applied current density $J$ via $J/J_0 = \kappa (1 - \kappa^2)$, and $J_d = (2/3\sqrt{3}) J_0 = (2/3\sqrt{3}) (\Phi_0/2\pi \mu_0 \lambda^2 \xi)$ is the GL depairing current density, $\rho_0$ the condensation energy density, and $S$ the cross section of the wire.
The prefactor in Eq.~\eqref{eq:rate} has been estimated by McCumber and Halperin to~\cite{McCumber1970b}
\begin{equation}
    \frac{\Omega}{2\pi} \approx \frac{\sqrt{3}}{2 \pi^{3/2} \tau_{GL} \xi} \sqrt{ \beta \Delta U(0) }(1 - \kappa \sqrt 3)(1 + \kappa^2/4) ,
\end{equation}
where $\tau_{GL} \propto |T - T_c|^{-1}$ is the GL time. The resulting rate is plotted in Fig.~\ref{fig:TDGL-rate}.
As before, the downturn at high currents is an artefact of the approximations.

\subsection{Vortex barrier crossing}
\label{sec:vortex-crossing}

For  wider superconducting 2D sheets or strips the switching transition will necessarily involve vortex crossings perpendicular to the current.
It is then natural to take the reaction coordinate $x$ to be the distance from the edge to the vortex center.
The applied sheet-current density $\mathbf J$ exerts a Lorentz force $\mathbf{f} = \mathbf{J} \times \mathbf{n} \Phi_0$ on a vortex trying to pull it further into and across the strip ($\mathbf{n}$ here is the normal to the surface).
We assume that the vortices undergo diffusive motion in a potential with diffusion coefficient $D = k_B T/\alpha$, where $\alpha = \Phi_0^2 / 2\pi \xi^2 \rho_n$ is the Bardeen-Stephen vortex friction~\cite{Bardeen-Stephen1965} related to the normal state resistivity $\rho_n$.

\begin{figure}
    \centering
    \includegraphics[width=1\linewidth]{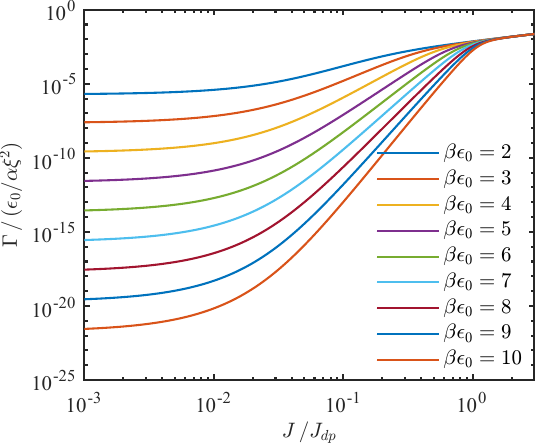}
    \caption{Switching rate $\Gamma$ vs.\ current computed from the inverse mean first passage time for the vortex barrier crossing. The width of the strip is $W = 100 \xi$. The different curves correspond to different temperatures.}
    \label{fig:vortex_rate}
\end{figure}

We further assume that the width $W$ of the strip is larger than the GL coherence length $\xi$ and much smaller than the Pearl length $\Lambda = \lambda^2 / d$, where $\lambda$ is the magnetic penetration depth and $d$ is the thickness of the strip.
The nucleation of a vortex at an edge (or an anti-vortex at the opposite edge) involves a depletion of the superconducting order parameter in a region of the order of $\pi\xi^2 d$, with an associated energy cost $\sim \epsilon_0/2$. As the vortex moves further into the strip, $\xi \lesssim x \lesssim W-\xi$, it will be attracted to its mirror images leading to an energy~\cite{Clem2011,Gurevich2008,Bulaevskii2011} $U_0(x) = \epsilon_0/2 + \epsilon_0 \ln [(2W /\pi\xi) \sin(\pi x/W)]$, where $\epsilon_0 = (\Phi_0^2 / 4\pi \mu_0 \lambda^2) d$.
A smooth interpolating formula for the total energy of the vortex may be defined as
\begin{equation}
    U(x,J) = \frac{\epsilon_0}{2} \ln \left[
    1 + e \left( \frac{2W}{\pi \xi}\right)^2 \sin^2 \left( \frac{\pi x}{W} \right)
    \right] - \Phi_0 J x.
\end{equation}
In Fig.~\ref{fig:vortex_rate} we plot the switching rate $\Gamma = \tau^{-1}$ obtained from the numerical solution to Eq.~\eqref{eq:MFPT} using this $U(x,J)$ for a couple of different temperatures and a width $W = 100\xi$.
Three different regimes are clearly seen:
For small currents $I = JW \ll \epsilon_0/\Phi_0$ the barrier maximum will occur near the center $x \approx W/2$ of the strip resulting in an energy barrier $U(J) \approx U_0 - \Phi_0 J W/2$.  The corresponding transition rate $\Gamma$ will then grow exponentially with current.
As the current is increased the maximum will move towards the edge and the dependence of the barrier on current turns logarithmic, which translates to a powerlaw dependence $\Gamma \sim J^b$, with an exponent
$b = \beta \epsilon_0 + 1$.
For larger currents corrections to the powerlaw behavior will occur due to the current-induced suppression of the order parameter not accounted for here \footnote{%
In GL theory a uniform current leads to the renormalization $\epsilon_0 \to \epsilon = (1-\kappa^2) \epsilon_0 < \epsilon_0$, where $\kappa$ is defined below Eq.~\eqref{eq:LAMH}}.
For even larger currents $J \gtrsim J_{dp} = (2/3\sqrt 3) (2 \epsilon_0/\Phi_0 \xi)$ the barrier will diminish resulting in yet another crossover to a regime where $\Gamma \sim J$.

\begin{figure}
    \centering
    \includegraphics[width=1\linewidth]{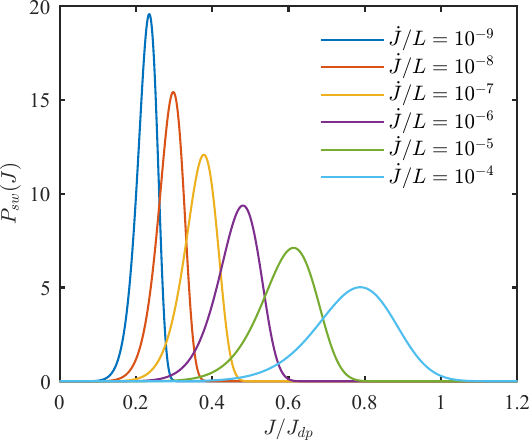}
    \caption{Example of switching distributions for the vortex barrier crossing scenario, for different values of sweep rate $\dot J$ and length $L$. The temperature is $k_BT = \epsilon_0/8$, $L = 100$ in units of $\xi$, and $\dot J$ is measured in units of $J_{dp} /(\xi^2/D)$.}
    \label{fig:vortex_switching}
\end{figure}

Correspondingly, the shape of the switching current distribution will change depending on what current regime is probed, which in turn depends on the sweep rate $\dot J$, wire length, and temperature.
For small currents the switching current will follow a Gumbel distribution, then a Weibull distribution with shape parameter $\beta \epsilon_0 + 2$ for higher currents, and eventually a Rayleigh distribution for the highest.
We show some examples of the switching distributions for various $\dot J / L$ and $\beta \epsilon_0 = 8$ in Fig.~\ref{fig:vortex_switching}.

\section{Time-dependent Ginzburg-Landau simulations}
\label{sec:TDGL}

A more detailed picture of the vortex barrier crossing requires a microscopic model including both amplitude and phase of the superconducting order parameter.
Therefore, we now turn to a larger scale numerical simulation of a 2D superconducting strip using a stochastic formulation of the time-dependent Ginzburg-Landau (TDGL) equations
\begin{equation}
    u\dot{\psi} = (1 - T - |\psi|^2)\psi + (\nabla - i \mathbf{A})^2\psi + \eta_\psi
\end{equation}
\begin{equation}
    \dot{\mathbf{A}} = Im\{ \psi^* (\nabla - i \mathbf{A}) \psi \} - \left( \frac{\lambda}{\xi} \right)^2 \nabla \times \mathbf{B} + \eta_\mathbf{A}
\end{equation}
\begin{equation}
    \mathbf{B} = \nabla \times \mathbf{A}
\end{equation}
here expressed in dimensionless units. Time is measured in units of the timescale $\tau_A = \mu_0 \sigma \lambda^2$, $\sigma$ is the normal state conductivity, and $u=\tau_{\psi} / \tau_A$ the ratio of the timescale of the two equations. Positions are in units of the coherence length $\xi$, $\lambda$ is the magnetic penetration length, $T$ is in units of the critical temperature $T_c$, and magnetic field measured in units of the $\hbar / 2 e \xi^2$.
The precise value of $u$ does not significantly influence the breakdown dynamics, and
for computational efficiency we set $u=1$ rather than the commonly used value $5.79$~\cite{Kramer1978TheoryFilaments} for a dirty superconductor.
The temperature is taken to be stationary and Joule heating effects due to dissipation and vortex motion have been neglected, however the stochastic dynamics associated with a finite temperature are included through white noise terms $\eta_\psi$ and $\eta_A$. 
The correlation functions of these noise terms are $\left< \eta_{\psi}(\mathbf{r}, t) \eta^{*}_{\psi}(\mathbf{r}', t') \right> = 4 u D T \delta(t - t') \delta(\mathbf{r} - \mathbf{r}')$ and $\left< \eta_{A}(\mathbf{r}, t) \eta_{A}(\mathbf{r}', t') \right> = 2 D T \delta(t - t') \delta(\mathbf{r} - \mathbf{r}')$, with the dimensionless coefficient $D = 2 e^2 \mu_0 k_B T_c \lambda^2 / (\hbar^2 d )$.
The thickness is here taken as $d=\xi$, and the critical temperature used is $T_c = 10$ K which is a typical order of magnitude for thin films of NbN or NbTiN used in detector devices \cite{Zichi19,SteinhauerYang2020,Sidorova2020}.
These equations are expressed in the zero electric potential gauge.

The geometry used to describe the wire is a rectangular domain $0<x<L$ and $0<y<W$ with length $L$ and width $W$. At $x=0$ and $x=L$ we use periodic boundary condition for both $\psi$ and $B$ to approximate the dynamics of a very long wire.
At $y=0$ and $y=W$ we apply the condition $\mathbf{n}\cdot (\nabla - i \mathbf{A})\psi = 0$ corresponding to an insulating boundary, where $\mathbf{n} = \pm \hat{\mathbf{y}}$ is the unit normal of the boundary.
The net current $I_\mathrm{net}$ through the cross section is tuned by the boundary conditions
$\mathbf{B}(x, 0) = - B_0 \hat{\mathbf{z}}$, $\mathbf{B}(x, W) = + B_0 \hat{\mathbf{z}}$
for the magnetic field, since
\begin{align}
    \frac{\mu_0 I_{\mathrm{net}}}{d} &= \int\limits_0^W (\nabla \times \mathbf{B}) \cdot \hat{\mathbf{x}} dy = 2 B_0.
\end{align}
For the numerical solution of the equations of motion we use an explicit finite difference scheme with fixed time step $\Delta t = 10^{-4}$, and the same spatial discretization $\Delta x = 1/2$ in both the $x$ and $y$ coordinates.
The complex phase $\theta$ and the vector potential $\mathbf{A}$ are invariant under the gauge transformation $\theta \rightarrow \theta + \alpha$, $\mathbf{A} \rightarrow \mathbf{A} + \nabla \alpha$. 
In order to preserve this gauge invariance of the covariant derivative we use a link-variable formulation $(\nabla -i \mathbf{A})\psi = U^* \nabla(U \psi)$, where $U = \exp(-i \int \mathbf{A}\cdot d\mathbf{l})$ is the so called link variable \cite{William1996}. This formulation of the derivative has the benefit of being explicitly gauge invariant when using the discretized finite difference approximation for the derivatives. 

In order to determine the switching event the bias current is increased continuously in time as $I(t) = \dot{I} t $ with a constant sweep rate $\dot{I}$, expressed in units of $\hbar d/(2 e \tau_A \mu_0 \xi^2)$. The switching current is identified from the time at which the breakdown of the superconducting state is first detected. This breakdown can be identified either by measuring the voltage along the strip, or alternatively by the flow of vortices across the wire.

In the zero-potential gauge the electric field is given by $\mathbf{E} = -\dot{\mathbf{A}}$ and the voltage along the strip is calculated through integrating the electric field
\begin{equation}\label{eq:line_voltage_integral}
    V = \int \dot{\mathbf{A}} \cdot d\mathbf{l}
\end{equation}
along a path between a point on left and the right hand boundaries of the strip. 
With the stochastic dynamics this voltage becomes a very noisy signal, and it is therefore inconvenient to use the condition $|V(I_c)| > 0$ to define the critical current $I_c$. 

The vortex flow across the wire can be identified by integrating the complex phase gradient along a path between the left and right hand boundaries
\begin{equation}\label{eq:line_winding_integral}
    \nu = \int \nabla \theta \cdot d\mathbf{l}.
\end{equation}
The path is taken to be the mid-line of the strip at $y=W/2$ between $x=0$ and $x=L$.
Each passing vortex increases this phase winding $\nu$ by $(\pm) 2\pi$, and the switching current is therefore determined as the first current for which $|\nu| \geq 2 \pi$.
In practice, for high enough bias current, the first vortex passage will trigger an avalanche of many more vortices flowing across the wire.
The first vortex passage therefore gives a more precise definition of the switching current compared to the detection of a non-zero voltage.

\begin{figure}
    \centering
    \includegraphics[width=1\linewidth]{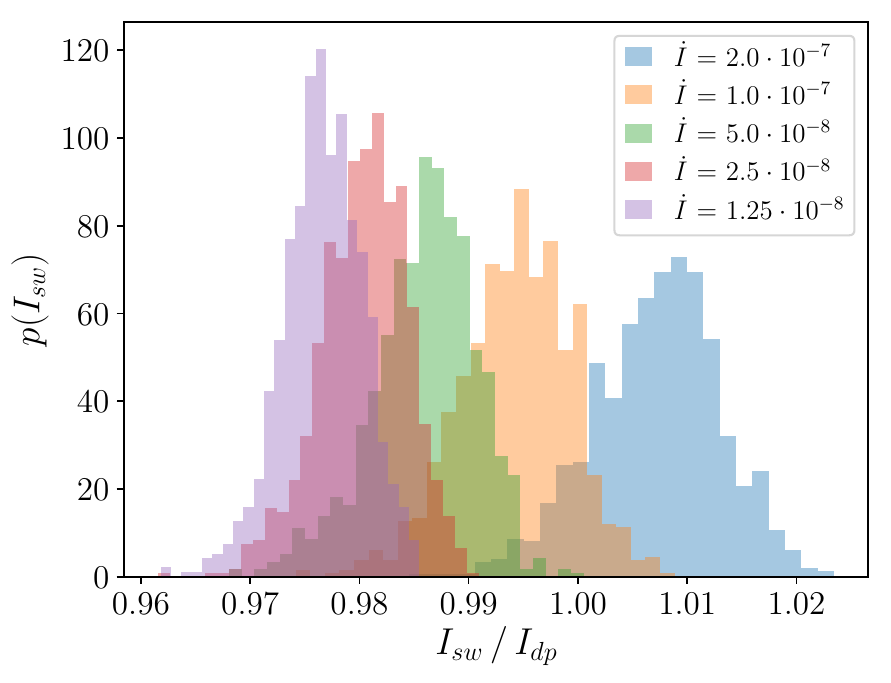}
    \caption{Switching current histograms computed from 1000 independent realizations of the stochastic dynamics for each value of the sweep rate $\dot{I}$ for a wire width $W=10\xi$, length $L=50\xi$ and $T/T_c = 0.8$.} 
    \label{fig:tdgl_ic_hist_sample}
\end{figure}

\begin{figure}
    \centering
    \includegraphics[width=1\linewidth]{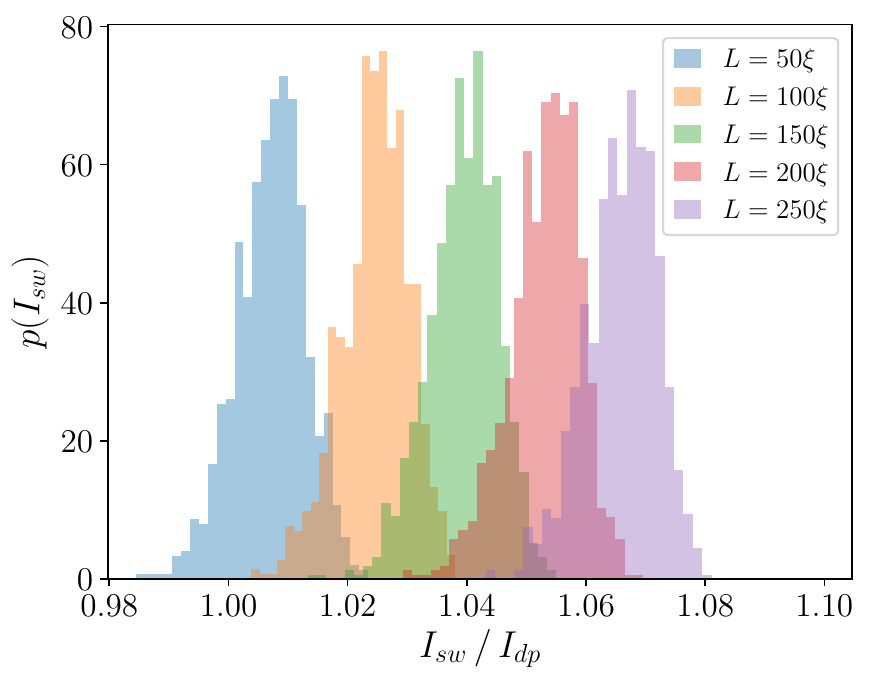}
    \caption{Switching current distributions computed for a wire width $W=10 \xi$ and ratio $\dot{I}/L = \mathrm{constant}$. The blue ($L=50\xi$) dataset shown here is the same as the blue dataset ($\dot{I}=2\cdot10^{-7}$) in Fig.~\ref{fig:tdgl_ic_hist_sample}.}
    \label{fig:tdgl_ic_hist_fixed_ratio}
\end{figure}

\begin{figure}
    \centering
    \includegraphics[width=1\linewidth]{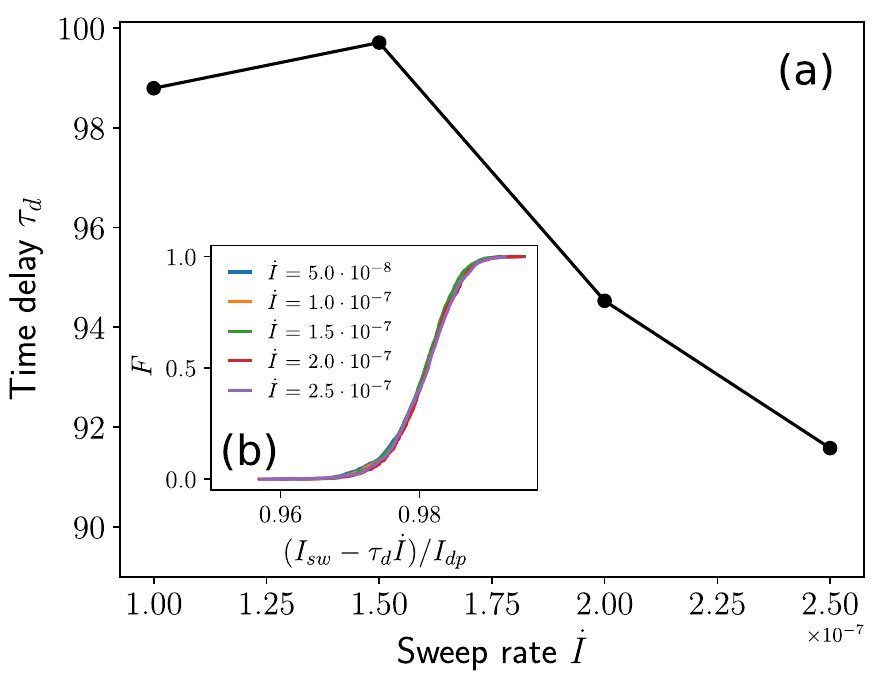}
    \caption{(a) The delay time $\tau_d$ in units of $\tau_A$ calculated with Eq.~\eqref{eq:tau_d_from_median} for the switching current distributions with fixed ratio $\dot{I}/L = \mathrm{constant}$ in Fig.~\ref{fig:tdgl_ic_hist_fixed_ratio}. The inset (b) shows the corresponding cumulative distributions $F$ shifted by the calculated value of the delay time $\tau_d$ for each sweep rate $\dot{I}$. }
    \label{fig:tdgl_delay_time_sample}
\end{figure}

\begin{figure}
    \centering
    \includegraphics[width=1\linewidth]{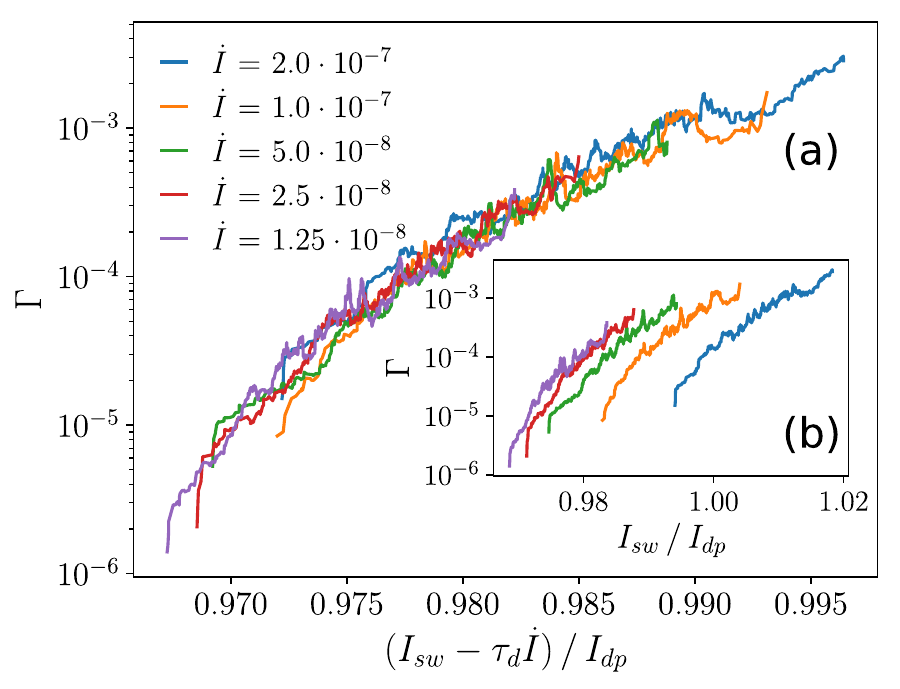}
    \caption{ (a) Switching rate $\Gamma$ expressed in units of $1/(\xi \tau_A)$ computed from the switching current distributions presented in Fig.~\ref{fig:tdgl_ic_hist_sample} according to Eq.~\eqref{eq:Gamma_from_discrete_derivative} for different sweep rates $\dot{I}$ and corrected by the time shift $\tau_d \approx 100 \tau_A$ computed for the slowest sweep rate shown in Fig.~\ref{fig:tdgl_delay_time_sample}.
    The inset (b) shows the corresponding $\Gamma$ obtained without accounting for this delay time, and it is clear that especially for fast sweep rates it is necessary to include $\tau_d > 0$ in order to obtain a significant overlap between the different curves for $\Gamma$.}
    \label{fig:tdgl_gamma_sample}
\end{figure}

\section{Simulation results}

Using the stochastic TDGL model we obtain statistics of the switching current for a wire with length $L$, width $W$, temperature $T$ for a sweep rate $\dot{I}$. The TDGL model allows us to simulate the vortex dynamics, and to define the switching current in terms of the first vortex passage. From switching current statistics we can therefore extract information about the vortex entry barrier, following the procedure described in Sec.~\ref{sec:switching-current-distr}.
An example of the switching current distribution of a relatively narrow wire ($W=10\xi$) is shown in Fig.~\ref{fig:tdgl_ic_hist_sample}, calculated from 1000 independent realizations of the stochastic noise for each $\dot{I}$. A large penetration length $\lambda = 20\xi$ is used to allow an approximately uniform distribution of the current density.

For these parameters the switching in all cases occur very close to the depairing current $I_{dp}$.
We can note that for faster sweep rates part of the distribution extends to $I_{sw} / I_{dp} > 1$. This is a consequence of a finite delay time $\tau_d$ between the initiation of the breakdown process and the detection of the first vortex at the midpoint $y=W/2$, as discussed in Sec.~\ref{sec:delay}. With a constant sweep rate $\dot{I}$ of the current this leads to an overestimation $\tau_d \dot{I}$ of the switching current.

In order to investigate this delay time numerically, we perform simulations for different $\dot{I}$ and with a fixed ratio $\dot{I}/L$, since according to Eq.~\eqref{eq:cummulative} the distribution $F(I_{sw})$ would be invariant in the absence of a delay time. The result is shown in Fig.~\ref{fig:tdgl_ic_hist_fixed_ratio}, 
and the fact that the histograms do not overlap while the overall shape of the distributions does not change is a clear indication that there is a significant time delay $\tau_d > 0$. 

While the $\tau_d$ in principle could depend on $\dot{I}$ we see from the figure that the histograms are separated by an approximately constant offset, indicating that $\tau_d$ is approximately constant. In order to estimate this time we make the ansatz $\bar{I}_i = \bar{I}_{sw} + \tau_d \dot{I}_i$, where $\bar{I}_i$ refers to the median of the distribution of the detected switching currents and $\bar{I}_{sw}$ the true median in absence of a time delay.
For each pair of distributions the $\tau_d$ can be obtained as
\begin{equation}\label{eq:tau_d_from_median}
    \tau_d = \frac{\bar{I}_2 - \bar{I}_1}{\dot{I}_2 - \dot{I}_1}.
\end{equation}
In order to evaluate this we let $\dot{I}_1$ be the slowest of the sweep rates, since this would be the least affected by the delay, and vary $\dot{I}_2$. The resulting delay time $\tau_d(\dot{I}_2)$ is shown in Fig.~\ref{fig:tdgl_delay_time_sample}(a).
For the slower sweep rates this delay approximately saturates to a constant value $\tau_d \approx 100 \tau_A$, however for larger $\dot{I}$ there is a small trend of gradually decreasing delay. This trend is interpreted as the larger $\dot{I}$ allowing the bias current to increase more before the first vortex has had time to nucleate, and the higher bias current in turn accelerating the nucleation process of the vortex. 
The inset in Fig.~\ref{fig:tdgl_delay_time_sample} (b) shows the cumulative distribution $F(I_{sw} - \tau_d \dot{I})$, 
where the detected switching current is corrected for the delay time $\tau_d$ calculated for each sweep rate individually.
We find that this shift is sufficient for the different distributions to collapse onto a single curve, and the obtained distributions $F$ display the expected dependence on the ratio $\dot{I}/L$ consistent with Eq.~\eqref{eq:cummulative}.

With the empirical cumulative distribution $F$ corrected for the delay time $\tau_d$ the switching rate $\Gamma$ can be calculated according to Eq.~\eqref{eq:Gamma_from_cumulative_dist}.
The derivative in this expression must be numerically evaluated on the set of switching currents $I_{sw}^{(i)}$ obtained for the $i=0,1,...,999$ independent simulation realizations.
The cumulative distribution is estimated as $F(I_{sw}^{(i)}) = i/1000$, where the sampled switching currents $I_{sw}^{(i)}$ are sorted in increasing order.
We use a symmetric difference approximation for the derivative as
\begin{equation}\label{eq:Gamma_from_discrete_derivative}
    \Gamma(I_{sw}^{(i)}) = - \frac{\dot{I}}{L} \frac{1}{I_{sw}^{(i+k)} - I_{sw}^{(i-k)}}\log\left( \frac{1 - F(I_{sw}^{(i+k)})}{1 - F(I_{sw}^{(i-k)})} \right).
\end{equation}
Using nearest neighbour differences ($k=1$) is found to result in a very noisy approximation of $\Gamma$, while similarly a very large value of $k$ would instead lead to a systematic underestimation of the variability of $\Gamma$, and we find a good middle ground in $k=20$. 

The resulting estimate of $\Gamma$ is shown in Fig.~\ref{fig:tdgl_gamma_sample}(a) for a fixed length $L=50\xi$ of the wire and different sweep rates $\dot{I}$ for the bias current. 
The switching currents are here corrected for the finite delay time by subtracting $\tau_d \dot{I}$, where we use the approximately constant value $\tau_d \approx 100 \tau_A$ obtained for the slower sweep rates in Fig.~\ref{fig:tdgl_delay_time_sample}.
The inset Fig.~\ref{fig:tdgl_gamma_sample} (b) for reference shows the corresponding $\Gamma$ where the switching current is not corrected by the shift $\tau_d \dot{I}$. It is clear that a finite $\tau_d$ must be taken into account in order for $\Gamma$ calculated for different sweep rates to collapse onto the same curve.

\begin{figure}
    \centering
    \includegraphics[width=1\linewidth]{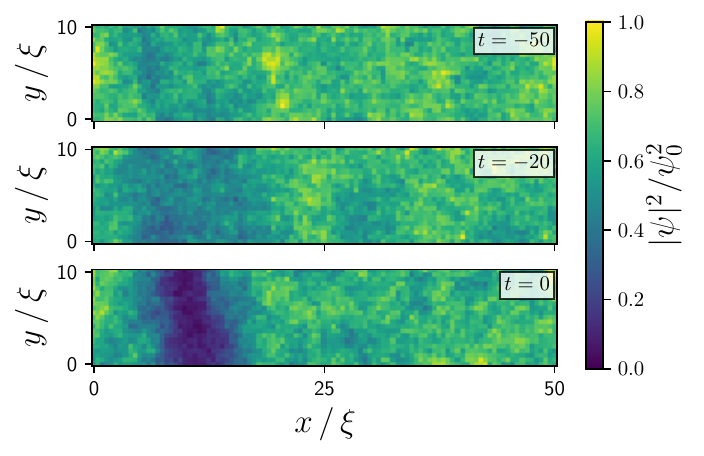}
    \caption{Snapshots of $|\psi|^2$ for different times $t$ during one realization of a switching event for narrow wire $W = 10\xi$, $\dot{I}=5\cdot 10^{-8}$, normalized by $\psi_0^2 = 1 - T$. The snapshot for time $t=0$ corresponds to the configuration when the first vortex is detected crossing the center line $y = W/2$, and a weakened domain is seen to slowly take shape for times $t < 0$. }
    \label{fig:tdgl_narrow_breakdown_snapshot}
\end{figure}

This nucleation time of a switching event can also be observed in snapshots of the modulus of the order parameter $|\psi|^2$. Examples of these snapshots are shown for a narrow wire ($W=10\xi$) in Fig.~\ref{fig:tdgl_narrow_breakdown_snapshot}, where $t=0$ corresponds to the time when the first vortex crossing is detected at the mid-line $y=W/2$, but a signature of the formation of a weakspot can be seen much earlier in this case at $t=-50 \tau_A$.

An additional contribution to the delay time is expected due to a finite vortex velocity~\footnote{A very rough theoretical estimate for the vortex motion contribution would be $W/v$, where the vortex velcocity $v \approx J \Phi_0 / \alpha$, neglecting any variation in the potential across the wire.}.
The size of this contribution can be estimated by detecting the first vortex crossing at several points $y_i$ along the cross section. 
For a vortex entering through the boundary $y=0$ and moving with a constant velocity $v$ the switching current profile would be of the form $I_{sw}(y_i) = I_{sw}(0) + \frac{y_i}{v} \dot{I}$. The $I_{sw}(0)$ here is the current at which the vortex first enters the strip, and the velocity $v$ can be estimated by fitting the slope.

\begin{figure}
    \centering
    \includegraphics[width=1\linewidth]{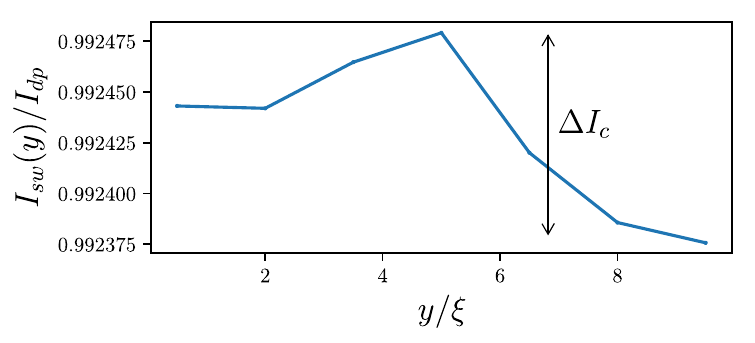}
    \caption{Switching current extracted at different points $y$ along the cross section for the realization shown in Fig.~\ref{fig:tdgl_narrow_breakdown_snapshot}. The variation $\Delta I_c$ is, however, small compared to $\tau_d \dot{I}$ computed in Fig.~\ref{fig:tdgl_delay_time_sample}.}
    \label{fig:tdgl_narrow_cross_section_ic}
\end{figure}

Often the breakdown will be more complex than a single vortex crossing from one boundary to the other, as an example of $I_{sw}(y_i)$ shows in Fig.~\ref{fig:tdgl_narrow_cross_section_ic}. The maximum near the center indicates that in this realization a vortex first entered at $y=W$, a short time later an anti-vortex enter at $y=0$ and the pair annihilates in the center.
The timescale associated with this process is estimated from the variation $\Delta I_c$ as $\Delta t_v = \Delta I_c / \dot{I}$ and is found to be much smaller than the timescale $\tau_d$ associated with the nucleation time of the first vortex described above.
Together Figs.~\ref{fig:tdgl_narrow_breakdown_snapshot} and \ref{fig:tdgl_narrow_cross_section_ic} suggest that, for a narrow ($W = 10\xi$) strip, a channel of suppressed order parameter amplitude forms prior to the passage of the vortices.

\begin{figure}
    \centering
    \includegraphics[width=1\linewidth]{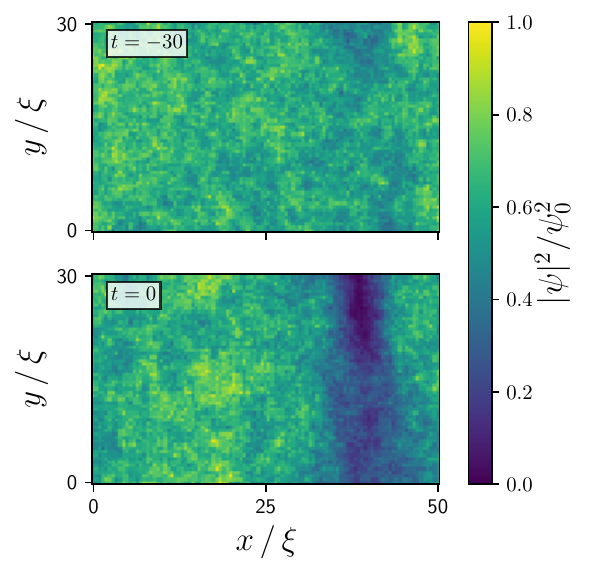}
    \caption{Snapshots of $|\psi|^2$ for different times $t$ during one realization of a switching event for wider wire $W = 30\xi$, $\dot{I}=2\cdot 10^{-7}$, normalized by $\psi_0^2 = 1 - T$. With a wider wire an elongated vortex can be seen in the snapshot.}
    \label{fig:tdgl_wider_breakdown_snapshot}
\end{figure}

\begin{figure}
    \centering
    \includegraphics[width=1\linewidth]{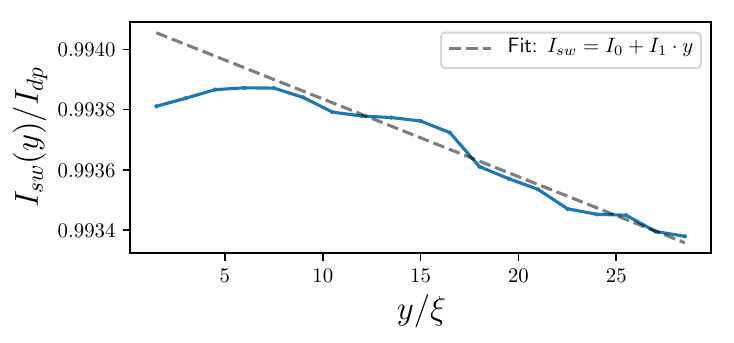}
    \caption{Switching current extracted at different points along the cross section for the realization shown in Fig.~\ref{fig:tdgl_wider_breakdown_snapshot}. An estimate for the speed of the vortex crossing can be extracted from fit, $v = |\dot{I}/I_1|$. }
    \label{fig:tdgl_wider_cross_section_ic}
\end{figure}

For a wider wire ($W=30\xi$) snapshots of the modulus $|\psi|^2$ show, Fig.~\ref{fig:tdgl_wider_breakdown_snapshot}, the same characteristic signature as seen in the narrow wire ($W=10\xi$), where a weak spot starts to form a relatively long time prior to the first vortex entry. In the wider wires it is possible to resolve the shape of the moving vortices. The influence of the driving current is seen to deform the vortex profile into an elongated shape. 
With a wider cross section it is natural that also the time delay due to the finite vortex velocity increases as shown in Fig.~\ref{fig:tdgl_wider_cross_section_ic}. For $W=30\xi$ this contribution to the total delay time $\tau_d$ is still small. However, as the width is increased further it will eventually become non-negligible.

\section{Discussion}

The simulated switching current distributions, shown in Fig.~\ref{fig:JJ} for a Josephson junction chain and in Fig.~\ref{fig:tdgl_ic_hist_sample} for a nanostrip,  have a characteristic asymmetric shape with more weight toward lower currents and standard deviation on the order of a few percent of the mean switching current. These distributions capture many of the same features seen in experimentally measured switching current distributions \cite{Ejrnaes2019,Parlato2020}.
The width of the simulated distributions decreases with decreasing sweep rate $\dot{I}$ of the bias current, which is mostly in agreement with experiments.
However, non-monotonous width-dependence has also been reported with a local minimum width appearing for intermediate values of $\dot I$~\cite{Baumans2017}.

The logarithm of the switching rate $\Gamma$ extracted from the simulated distributions 
follows a slightly concave curve, see Fig.~\ref{fig:tdgl_gamma_sample}.
A perfectly straight line, i.e., exponential dependence of $\Gamma$ on current, would correspond to a Gumbel distribution for the switching current, which has a skewness of $-1.14$.
The concave dependence we see correspond to slightly less skewed distributions, e.g., with skewness about $-0.5$ for the TDGL simulations.
This concave trend of the switching rate has also been seen in experiments~\cite{Ejrnaes2019,Parlato2020}, with increasing curvature for higher temperatures. In these references the curvature was attributed to multiple phase-slips being required in order to trigger a full switching event. However in our simulations the switching event is determined by the first phase-slippage detected, which is an indication that the curvature alone may not be a unique signature of a multi-phase-slip regime.
In fact, under the assumptions discussed in Sec.~\ref{sec:theory} the rate $\Gamma(I)$ reflects the current dependence of the energy barrier $\Delta U(I)$, which may be complicated with several different crossovers.

In our TDGL simulations of narrow wires we observed that the vortex nucleation process begins with a growing depletion of the superconducting order parameter amplitude at one edge of the strip. Only once the order parameter is sufficiently suppressed over a region reaching across the width of the strip do vortices start to flow.
When the width is increased it becomes possible to discern the flow of individual vortices.
Eventually, for even wider strips the vortex crossing scenario discussed in Sec.~\ref{sec:vortex-crossing} should become applicable.

By exploiting how a slower sweep rate $\dot{I}$ shift the $I_{sw}$ distribution towards lower currents, the result for different values of $\dot{I}$ can be combined according to Eq.~\eqref{eq:Gamma_from_cumulative_dist}. This permits extracting the switching rate $\Gamma$, and by extension the vortex energy barrier $\Delta U(I)$, on a larger interval of currents.
In doing this for our simulation results we found it necessary account for a time delay $\tau_d$ between the initiation of a phase slip and its later detection in the form of a vortex crossing.
This intrinsic time delay is estimated to be of the order of a few ps in superconducting materials such as NbN using typical values for $\sigma$ and $\lambda$ \cite{Bartolf2010},
making it less of an issue in experimental settings.
On the other hand, experiments may be subjected to other more important contributions to the delay time, such as propagation time of the voltage pulse through the wire, that may need to be taken into account.

By lowering the sweep rate or equivalently studying longer wires it is possible to probe rare barrier crossings.
In this regime it is likely that inhomogeneities and disorder will start to play a role.
Studying this crossover to a disorder dominated regime~\cite{Chen2024} would be an interesting extension.
Several experiments have also demonstrated a non-monotoneous temperature dependence of the width of the switching current distribution~\cite{Murphy2015,Ejrnaes2019,Parlato2020}. Such an effect has been attributed to Joule-heating \cite{Shah2008}, which has not been included in this investigation but could be another potential avenue for a future work.

\section{Conclusion}

The formula \eqref{eq:Gamma_from_cumulative_dist} connects the switching rate $\Gamma$ to the probability distribution of switching currents $P(I_{sw})$.
This formula is particularly useful for analyzing simulation data as we demonstrate above, and also for experiments.
A complication is that in practice the switching current histograms may need to be shifted due to a prevalent time delay. Accounting for this, our simulations fit well with the theoretical description in Sec.~\ref{sec:switching-current-distr} and makes it possible to stitch together the extracted switching rate $\Gamma$ from several switching current distributions taken using different sweep rates $\dot I$.
The rate $\Gamma$ in turn, makes it possible to obtain the thermal activation energy barrier of the process to logarithmic accuracy, i.e., neglecting the prefactor in Eq.~\eqref{eq:rate}, thus providing insight into the detailed switching mechanism.
A similar analysis could be employed for experimental data, thus allowing both the switching rate $\Gamma$ and time delay $\tau_d$ to be measured.

\begin{acknowledgments}
The computations were enabled by resources provided by the National Academic Infrastructure for Supercomputing in Sweden (NAISS), partially funded by the Swedish Research Council through grant agreement no. 2022-06725.
\end{acknowledgments}



\bibliography{switch,robert_refs}

\end{document}